# Structure and Magnetism in the Bond Frustrated Spinel, ZnCr$_2$Se$_4$

P. Zajdel[1], W-Y. Li[2], W. Van Beek[3], A. Lappas[4], A. Ziolkowska[5], S. Jaskiewicz[5], C. Stock[5] and M. A. Green[6,*]

[1] *Institute of Physics, University of Silesia, ul. Uniwersytecka 4, 40007, Katowice, Poland.*

[2] *Department of Chemistry, University College London, Gordon Street, London, WC1H 0AJ, UK.*

[3] *European Synchrotron Radiation Facility, Polygone Scientifique Louis Néel, 6, Rue Jules Horowitz, 38000 Grenoble, France.*

[4] *Institute of Electronic Structure and Laser, Foundation for Research and Technology – Hellas, 71 110 Heraklion, Greece.*

[5] *School of Physics and Astronomy, University of Edinburgh, Edinburgh, EH9 3JZ, UK.*

[6] *School of Physical Sciences, Ingram Building, University of Kent, Canterbury, Kent CT2 7NH, UK.*

* m.green@kent.ac.uk

The crystal and magnetic structures of stoichiometric ZnCr$_2$Se$_4$ have been investigated using synchrotron X-ray and neutron powder diffraction, muon spin relaxation (µSR) and inelastic neutron scattering. Synchrotron X-ray diffraction shows a spin-lattice distortion from the cubic $Fd\bar{3}m$ spinel to a tetragonal $I4_1/amd$ lattice below $T_N$ = 21 K, where powder neutron diffraction confirms the formation of a helical magnetic structure with magnetic moment of 3.04(3) µ$_B$ at 1.5 K; close to that expected for high-spin Cr$^{3+}$. µSR measurements show prominent local spin correlations that are established at temperatures considerably higher (< 100 K) than the onset of long range magnetic order. The stretched exponential nature of the relaxation in the local spin correlation regime suggests a wide distribution of depolarizing fields. Below $T_N$, unusually fast (> 100 µs$^{-1}$) muon relaxation rates are suggestive of rapid site hopping of the muons in static field. Inelastic neutron scattering measurements show a gapless mode at an incommensurate propagation vector of $k$ = (0 0 0.4648(2)) in the low temperature magnetic ordered phase that extends to 0.8 meV. The dispersion is modelled by a two parameter Hamiltonian, containing ferromagnetic nearest neighbor and antiferromagnetic next nearest neighbor interactions with a $J_{nnn}/J_{nn}$ = -0.337.

## I. Introduction

ZnCr$_2$Se$_4$ has been the subject of numerous studies in the past few decades since being reported as a canonical incommensurate simple helical magnet[1]. Its p-type semiconductor ($E_g$ = 0.3 eV)[2] band structure contrasts with the more common itinerant ferromagnetic helical structures, such as that found in elemental Ho[3] or MnSi[4]. Such complex magnetic systems have recently attracted renewed attention due to their potential route to colossal magnetoresistance[5] and multiferroic properties[6,7]. More recently, the importance of incommensurate magnetic spiral structures have been highlighted in the field of iron based superconductors and related systems[8-14]. ZnCr$_2$Se$_4$ displays both significant spin-phonon coupling[15] and magnetostriction[16], demonstrating the importance of fully understanding its structure and magnetism.

ZnCr$_2$Se$_4$ has a large positive Curie-Weiss temperature of $\theta_{CW}$ = 110 K[17], as a result of strong near 90º Cr–Se–Cr ferromagnetic superexchange, so does not show the geometric frustration present in many spinels. However, bond frustration reduces the magnetic ordering temperature to 21 K[1], which is significantly below $\theta_{CW}$, and is antiferromagnetic as a result of the collective strength of next nearest neighbor interactions[2,18]. Theoretical work has demonstrated that the stability of the helical structure in ZnCr$_2$Se$_4$ relays on the interplay of five magnetic exchange pathways[19], including the two large nearest neighbor ferromagnetic Cr-Se-Cr superexchange terms (∠~90°, $J$ ~ 25.4 K), and next nearest neighbor Cr-Se-Cr-Se-Cr and Cr-Se-Zn-Se-Cr ($J'$ ~ -6.6 K) antiferromagnetic exchange, which has been experimentally corroborated[20]. The extent of the bond frustration in ZnCr$_2$Se$_4$ ($\theta_{CW}/T_N$ ~ 5.2) is high even compared with other spinel structures. Spin fluctuations in ZnCr$_2$Se$_4$ were observed by diffuse magnetic scattering above $T_N$ suggesting a magnetic microdomain phase below 45 K[21,22]. Further studies conducted between the Curie-Weiss and Neel temperatures, revealed the presence of negative thermal expansion below 100 K[16,23], shifting of the magnetic resonance line[20], anomalous behavior of eigenfrequencies of phonon modes[16] and deviation from Curie-Weiss behavior[1], making further investigation with muon spin relaxation measurements attractive. The presence of local magnetic correlations is well established in the analogous oxide, ZnCr$_2$O$_4$[24-26], but which possesses, in contrast to selenide, a large negative Curie-Weiss constant. Therefore, the suppression of long-range magnetic order in ZnCr$_2$O$_4$ ($\theta_{CW}/T_N$ = 32) is the result of inherent geometric frustration and not only from bond frustration. The ability to probe local field distribution in time and space makes muon spin relaxation[27] an ideal technique to investigate the local magnetism in $ZnCr_2Se_4$. For example, it was successfully employed in the investigation of the onset of local correlations above $T_N$ in $Li_2Mn_2O_4$[28].

Spinel compounds, such as $ACr_2O_4$ (A = Zn or Cd)[26,29,30], often adopt lower symmetry structures to relieve magnetic frustration. For $ZnCr_2Se_4$ the case is not clear as neutron diffraction results suggested an *Fddd* orthorhombic structure[31], whereas synchrotron diffraction did not reveal a distinct structural transition, but showed unusual trends in Zn-Se bond lengths that were indicative of some structural modifications[32]. Previous neutron studies have reported a significantly reduced moment on the Cr ions in the helical arrangement, suggesting substantial magnetic disorder[32,33], possibly coming from a frustrated lattice configuration. Here we show that $ZnCr_2Se_4$ adopts a tetragonal *I4$_1$/amd* nuclear structure below $T_N$ and the helical magnetic structure contains Cr moments of 3.04(4) $\mu_B$; close to the theoretical spin-only values for high-spin $Cr^{3+}$. These synchrotron and neutron scattering studies are complemented with local µSR magnetic studies and magnetic fluctuations through inelastic neutron scattering.

**II. Experimental Methods**

$ZnCr_2Se_4$ was synthesized by combination of the elements in an evacuated (< $10^{-5}$ torr) quartz tube and heated at 900°C for several days with intermittent regrindings. Neutron scattering measurements were performed at the NIST Center for Neutron Research, National Institute of Standards and Technology, Gaithersburg, USA. High resolution neutron powder diffraction was performed on the BT-1 diffractometer between 1.2 and 300 K. A Ge (311) monochromator at a 75° take-off angle with $\lambda$ = 2.0782 Å, were used to collect data over a 1.3 – 166.3° angular range with a step size of 0.05°. The magnetic excitations in $ZnCr_2Se_4$ were studied using the Disk Chopper Spectrometer (DCS) with $E_i$ = 3.3 meV and 1.7 meV. Long wavelength neutron powder diffraction data was extracted between -0.2 and 0.2 meV from the elastic contribution of the DCS data at $\lambda$ = 5.0 Å at 1.5 K. Chemical analysis was performed using Prompt Gamma Activation Analysis[34] on the NG-7 instrument[35] at NIST Center for Neutron Research. High resolution powder X-ray diffraction experiments were performed at both the Swiss-Norwegian and ID31 beamlines at the European Synchrotron Radiation Facility, Grenoble, France from 5 to 500 K, with wavelengths of $\lambda$ = 0.72323(1) and $\lambda$ = 0.40016(7) Å, respectively. Both powder neutron and X-ray diffraction datasets were refined using Fullprof program[36]. Muon spin relaxation measurements were performed on the MuSR instrument, at ISIS facility, Rutherford Appleton Lab,

UK, and the GPS/SµS beamline of the Paul Scherrer Institut, Villigen, Switzerland. Data were fitted using the WIMDA package (ISIS) and MUFIT program (PSI).

## III Results and Discussion

**Synthesis**

The synthesis temperature and heating times were found to be critical to the resultant composition of the structure, which also varied greatly with the target mass of the final product. Deviation of the stoichiometry is well established for these systems, as well as extensive changes to the resultant electronic properties[2]. In order to monitor the process and ensure the complete reaction of the constituent elements a variety of characterization techniques were employed. Samples that were heated significantly below 900ºC failed to react completely and showed poor crystallinity, whilst those heated significantly above this temperature proved not to achieve exact stoichiometry. The procedure that proved to be the most successful was to heat samples at 900ºC in an evacuated tube for 24 hour periods. Initial monitoring of the reaction was performed using laboratory X-ray powder diffraction. If unreacted starting materials were present, the samples were reground and returned to the furnace. Once the amount of unreacted elements dropped below the level that could be detected by X-ray diffraction then Rietveld refinement of powder neutron diffraction data using the BT1 diffractometer at NIST was performed after each heating procedure until exact composition was achieved. The final stoichiometry of the sample used in the present study was confirmed to be $Zn_{1.003(6)}Cr_{2.01(1)}Se_{3.998(7)}$ by a combination of powder neutron diffraction and Prompt Gamma Activation Analysis. Additional confirmation was obtained from Rietveld refinement of the synchrotron powder diffraction data at room temperature using ID31 diffractometer, which confirmed a composition of $Zn_{1.000(8)}Cr_{1.99(2)}Se_{4.01(3)}$. It was concluded that within the errors of these experiments the compound was exactly stoichiometric.

**Variable Temperature Powder Synchrotron X-ray Diffraction**

To perform the symmetry analysis on the neutron diffraction data and determine the precise magnetic structure, it is essential to determine the low temperature space group of the crystal structure. It is well established that strong spin-lattice effects lower the symmetry from the ideal $Fd\bar{3}m$ space group in many spinels. To evaluate the true symmetry below $T_N$, we performed high-resolution synchrotron X-ray diffraction measurements on ID31 at the ESRF. The possible maximal subgroups of $Fd\bar{3}m$ are:

141 $I4_1/amd$ tetragonal
166 $R\bar{3}m$ rhombohedral
203 $Fd\bar{3}$ cubic
210 $F4_132$ cubic
216 $F\bar{4}3m$ cubic

The data collected at 5 K indexed to a tetragonal $I4_1/amd$ cell and full Rietveld refinement using the Fullprof package, which is shown in Figure 1 (a), confirmed a distortion to a subgroup of $Fd\bar{3}m$. Attempts to refine the data in an orthorhombic symmetry including the recently proposed $Fddd$ symmetry[37,38] or other tetragonal symmetries led to poorer fits. A summary of the key tetragonal models is provided in Table 1. Final refined parameters at 50 K were lattice parameters of $a = 10.486154(5)$ Å, giving a cell volume of 1153.051(1) Å$^3$ and atomic coordinates of Zn (0.125 0.125 0.125), Cr (0.5 0.5 0.5) and Se (0.25960(1) 0.25960(1) 0.25960(1)). The refinement at 5 K the $I4_1/amd$ symmetry, gave lattice parameters of $a = 7.41947(1)$ Å and $c = 10.48653(1)$ Å which gives a cell volume of 577.267(1) Å$^3$ and atomic coordinates of Zn (0 0.25 0.375), Cr (0 0 0) and Se (0 0.48066(3) 0.25944(2)). A comparison of a section of the Rietveld refinement ~ 13° demonstrates the splitting of the [622] reflection in the cubic space group at 50 K, shown in Figure 1(b), to the [422] and [206] reflections in the tetragonal symmetry, shown in Figure 1 (c).

At 50 K in the cubic $Fd\bar{3}m$ space group, the Zn-Se is in a perfect tetrahedron possessing $\bar{4}3m$ site symmetry. Any value of the structural parameter, u, in the Se (u u u) position, above the ideal spinel value of 0.25 represents a movement of the Se anion down the [111] direction and a homogeneous expansion of the tetrahedron. In the case of $ZnCr_2Se_4$ at 50 K, the value of 0.25960(1) represents an expansion of the tetrahedron by ~0.175 Å so that all Zn – Se bond distances are 2.4447(3) Å. There is a consequential reduction in the Cr – Se bonds by ~0.1 Å, which reduces the site symmetry from octahedral $m\bar{3}m$ to a $\bar{3}m$ trigonal antiprism symmetry. Although all six Cr – Se bonds are equidistance at 2.5249(1) Å, there is a distortion to 6 Se – Cr – Se bond angles to 85.343(7) Å and 6 Se – Cr – Se bond angles to 94.657(7)Å, which would be 90° at u = 0.25.

Calculating the equivalent cubic lattice parameters of √2 x 7.41947(1) for $ZnCr_2Se_4$ at 5 K, gives a $c/a$ ratio of 0.99941. Figure 2(a) shows a comparison of the Zn – Se and Cr – Se bond lengths, along with the Se – Zn – Se bond angle, as a function of temperature from 5 to 50 K. Despite the significant magnetoelastic transition at $T_N$ that reduces the symmetry to the $I4_1/amd$ space group, there is no difference within experimental error to either the four Zn – Se or the six Cr – Se bond distances. Within the transition the change in the *c/a* ratio is entirely accommodated by a modification to the $CrSe_6$ trigonal anti-prism and consequently the $ZnSe_4$ units. This is best demonstrated by inspection of the Zn – Se bond angles as a function of temperature, which is shown in Figure 2 (a), where the ideal bond angle of 109.47° is distorted to two different values; at 5 K two are at 109.57(4) Å while the remaining four are at 109.42(2) Å. It is important to emphasise that despite this distortion and reduction in symmetry so that the two Cr - Se bonds along the c direction are inequivalent to the four Cr - Se bonds in the ab plane, they remain, within the resolution of the experiment the same length; at 5 K two are at 2.5253(8) Å compared with four at 2.5250(5) Å. The reduction of the Cr site symmetry to 2/m produces 4 different Se-Cr-Se bond angles, 2 x 85.33(2)°, 4 x 85.36(2)° and 2 x 94.67(2)°, 4 x 94.64(2)°. Further analysis shows that the greatest change between the two structures above and below $T_N$ is in the Cr thermal factor, which is shown in Figure 2 (b) along with the change in lattice parameters as a function of temperature. The large increase in the Cr thermal parameter is likely to be the result of a distribution of local static displacements, which are being averaged by the diffraction experiment, as a result of the variation of magnetic exchange each of the $Cr^{3+}$ ions experience through the incommensurate nature of the magnetism.

**Magnetic Structure**

With the knowledge of the correct space group, powder neutron diffraction was used to further evaluate the structure and magnetism at low temperature. Although previous work all agree on a simple helical magnetic structure, one of the key unresolved issues is the magnitude of the moment, which has been reported to be 1.7[33,39] and 1.9 $\mu_B$[32], all of which are well below that expected for $Cr^{3+}$. Rietveld refinement of the low temperature magnetic structure was performed using a magnetic cone in real space[40]. The agreement factors $R_{Bragg}$ for nuclear and $R_{mag}$ for magnetic structures were 1.9% and 9.6%, respectively. The refined value of the ordered magnetic moment on chromium was found to be 3.04(3) with a propagation vector of *k* = (0 0 0.4648(2)). The final fits to the neutron diffraction data are shown in Figure 3 (a), along with a graphical representation of the helical magnetic model in Figure 3 (b). The magnitude of the ordered moment is that expected for an $S = 3/2$ system, contrary to previous reports that have reported reduced moment[32,33,39].

**Local Magnetism from Muon Spin Relaxation**

Data from muon spin relaxation measurements, performed at both the ISIS facility, UK and Paul Scherrer Institute, Switzerland, were fitted with simple phenomenological models at temperatures below and above $T_N$. In the magnetically ordered phase below $T_N$ for zero field (ZF) measurements, we applied a combination of two functions. An exponentially damped oscillation that describes the initial dip in the muon asymmetry with a non-zero slope of initial relaxation coupled with a slow exponentially relaxing background, which overall takes on the form written in equation [1] below:

$$G_{below}(t) = A_1 exp(-\lambda_1 t)cos(2\pi v t + \varphi) + A_2 exp(-\lambda_2 t) \qquad [1]$$

Similar approach was previously employed in the frustrated magnet α-$Zr_{1-x}Fe_x$[41] and the significant non-zero slope of the relaxation at early times makes $exp(-\lambda t)cos(2\pi v t+\varphi)$ more suitable than the $exp(-\sigma^2 t^2)J_0(\omega t)$ factor with 0 derivative at $t \to 0$ used for the double spiral system $CeAl_2$[42].

In the temperature regime above $T_N$, we used a single stretched exponential decay in the form of equation [2] below [27,42-45]:

$$G_{above}(t) = A_0 exp[(-t\lambda_3)^k] \qquad [2]$$

In this case, the exponent $k$ is preferred instead of the usual term, $\beta$, to avoid confusion with the critical exponent.

The justification for the selected phenomenological models is presented below. There are a limited number of examples in the literature of µSR on true incommensurate helical magnetic structures, except for the double helix model in $CeAl_2$[42] and some numerical simulations for $Cr_{1-x}Mn_x$[46]. Much more effort has been placed on incommensurate spin density wave (SDW) materials[47], where the initial relaxation was successfully modeled using spherical Bessel $J_0$ function or the itinerant MnSi[48,49], which shows a Kubo-Toyabe type of relaxation. For numerous reasons, our $ZnCr_2Se_4$ data could not be fitted below the $T_N$ to any of these previously reported models. Firstly, the use of Bessel $J_0$ functions implies a zero relaxation at the t → 0 limit, whilst our data shows non-zero values (see Figure 4), although the presence of such fast relaxation might prohibit the measurement of a zero-slope region. Secondly, there is a sharp dip in the asymmetry (see 5 K data in Figure 4a) at early time period, which is impossible to fit using a Bessel function alone. Thirdly, we lack the presence of a substantial non-relaxing tail below $T_N$.

The theoretical models for local field distribution in the case of a spin density wave can be approximated by the Overhauser distribution, which is a continuous distribution of fields between 0 and a singularity at $B_{max}$[50,51]. In a helical incommensurate magnetic structure, a low field cut-off is additionally present[42,46], so the local field is never equal to zero. The dominant value of $B_{max}$ plays a role of the coherent local field, and generates the oscillatory like feature "the dip" in the depolarization rate, which is modeled using a cosine function in the first component on the Eq. 1. The maximum value of the precession frequency is below 40 MHz, which is equivalent to the internal field ($B_{max}$) of about 0.3 T. It is important to note that despite being incommensurate, the presence of ferromagnetic Cr planes creates a local magnetic field that is mostly two-dimensional in character. The inset in Figure 5 shows that the onset of the oscillation frequency at $T_N$ caused by the creation of long-ranged helical magnetic order is associated with the appearance of a fast depolarizing component, which damps the oscillations. This fast relaxation coupled to the oscillatory part is taken into account in the first component of Equation [1] using the exponential factor $exp(-\lambda_1 t)$. The relaxation at the long times is modeled using a second exponent $exp(-\lambda_2 t)$.

In order to explain the lack of the non-relaxing tail below $T_N$ in the ZF data, let us note that in both the case of α-$Zr_{1-x}Fe_x$[41] and $CeAl_2$[42], the importance of additional spatial field averaging is highlighted, which is caused by the fast muon hoping between symmetrically equivalent sites resulting in a distribution of muon frequencies. The application of a 0.3 T longitudinal field (LF) with the strength of the order of $B_{max}$ (see Figure 4b and 6) removes the initial dip and significantly reduces, but not fully removes, the depolarization from 150 to 20 μs$^{-1}$ below $T_N$, which is consistent with the static ordered moment being aligned in the direction of the field.

The stretched exponent form of relaxation (Eq. [2]) used above the $T_N$ was found to model the data better than sum of two exponents Eq. [1] with $v = 0$ Hz. The difference between a stretched exponent and a double exponent model has been previously evaluated in the case of manganates, where a simple procedure was suggested to differentiate between the two[52]. In such an evaluation, a comparison is made between the resulting curves obtained by plotting $-t/ln(G(t))$ against $ln\ t$, which, in both cases, are linear with a gradient of $1-k$. The measurements for $ZnCr_2Se_4$ at 0.3 T LF and 25 K indicate a power type of decay with exponent equal to 0.6, which was also the best fit obtained for the zero-field data above the point where $\lambda_2$ reaches its maximum. Therefore, the high temperature regime was subsequently fitted with a stretched exponential with variable exponent, instead of two exponentially damped components given in equation [1]. The results from the final fits are shown in Figures 5 and 6.

One of the most significant finding of the zero field (ZF) data above the magnetic transition, is the presence of a dramatic change in the relaxation rate around 100 K, which is shown in Figure 5. The relaxation rate $\lambda_3$ increases monotonically when approaching the $T_N$ from higher temperature, which is expected behavior due to the critical slowing down of spin fluctuations. Whereas the exponent $k$ shows a dome-like behaviour with temperature, possessing a value of ~ 0.5 at both $T_N$ and 100 K, whilst showing a maximum at a mid-temperature point at ~ 50 K.

Since the stretched exponential form of relaxation can be explained in the terms of multiple exponential curves, but with variable rates[44], it is reasonable to assume that $ZnCr_2Se_4$ starts to develop areas of correlated ferromagnetic spins at the temperatures above the 100 K. The development of a long-ranged order at its Curie-Weiss value of 110 K[17] is hindered by the bond frustration due to multiple competing interactions present in the lattice[19,20], leaving uncorrelated two dimensional ferromagnetic planes. The signature of this is found in the temperature range 50 K - 100 K, where the value of stretched exponent approaches 1, which indicates a unification of relaxation rates. As the application of a 0.3 T LF is not sufficient to fully align the Cr moments, the muons are still depolarized above $T_N$, but below 100K, and the power factor $k$ still retains slight anomalous temperature dependence, as shown in Figure 6.

**Inelastic Neutron Scattering**

At low temperatures of 2 K, shown in Figure 7 (a), in the magnetically ordered phase, a gapless mode originating from the magnetic ordering wavevector of (0 0 0.4684(4)) is seen with a higher energy mode extending up to at least 0.8 meV. At higher temperatures of 30 K, shown in Figure 7 (b), the low temperature structured spectrum is replaced by momentum and energy broadened fluctuations, whilst Figure 7 (c) illustrates a high temperature scan showing the background and the clear absence of any magnetic fluctuations.

Given the lack of an orbital degree of freedom in $Cr^{3+}$, we consider the origin of the spiral order as resulting from competing symmetric exchange terms in the Hamiltonian, and compare this against scattering and thermodynamic data. Such a description based on excitations and diffraction has been successfully applied to other insulating magnets to understand the origin of the incommensurate ordering wavevector[53,54]. Previous theoretical studies of spinels[19] have considered this symmetric exchange only approach and noted a number of exchange constants (at least 5 in the case of reference [19]) need to be considered given the local bonding in the ideal (undistorted) spinel framework.

To simplify this description and to provide a tractable means of parameterizing our powder data, we consider a heuristic model based upon coupled chains. We therefore write a Hamiltonian with a ferromagnetic nearest neighbor exchange term ($J_{nn}$) for coupling within a given $Cr^{3+}$ chain, as well as the coupling to nearest neighbor chains, which originates from the 90º Cr-Se-Cr bond. For simplicity, we approximate both the exchange constants within and between chains (inter and intra chain) to be equal given the bonding close to 90º. We then consider a second exchange coupling $Cr^{3+}$ spins in one chain to the next nearest neighbor chain, which we denote here as $J_{nnn}$. Given that this interaction is more complex, we consider this exchange term to be antiferromagnetic to stabilize spiral magnetic order. This approach is broadly consistent with the analysis of reference [19] where nearest neighbor exchange parameters were ferromagnetic, whereas further exchange was antiferromagnetic to stabilize the observed magnetic structure. This simplified heuristic model provides us a means of parametrizing the data in terms of only two exchange constants describing coupling between $Cr^{3+}$ chains rather than a more complex series, which is difficult to uniquely determine for powder data. As noted in reference [19] (and in particular Table II of that work) there are multiple possibilities of the 5 expected exchange constants that could result in the required spiral structure. The Hamiltonian in this simplified heuristic model takes the form,

$$H = -J_{nn} \sum_j \vec{S}_j \cdot \vec{S}_{j+ab} - \sum_i (J_{nn} \vec{S}_i \cdot \vec{S}_{i+1} + J_{nnn} \vec{S}_i \cdot \vec{S}_{i+2})  \quad [3]$$

We first consider the ordering wavevector of (0 0 0.4684(4)). The incommensurate ordering wave vector in our symmetric-only simplified magnetic Hamiltonian considering connectivity along *c* (in terms of *L*) can be obtained by minimizing the classical exchange energy derived from the above Hamiltonian,

$$E(L) = J_{nn} \cos(2\pi L/4) + J_{nnn} \cos(2\pi L/2)  \quad [4]$$

By minimizing this classic energy, we find that *E(L)* is minimized at *L* = 0.468 when $J_{nnn}/J_{nn}$ = -0.337. The ordering wavevector only gives the ratio of the nearest and next nearest neighbor exchange constants and not the absolute values. To obtain an estimate of the exchange constants we now investigate the spin excitations in a powder of ZnCr$_2$Se$_4$.

The powder averaged spin excitations are illustrated in Figures 7 and 8. At low temperatures the data shows a low-energy branch extending up to 0.5 meV, and a higher energy branch to around 0.8 meV. The momentum range spanned by the experiment covers the first Brillouin zone of ZnCr$_2$Se$_4$, and therefore we compare the data against single crystal dispersion relations. Within the symmetric only model above, the spin-wave dispersion is obtained using the following equation[55].

$$(\hbar\omega)^2 = \langle S \rangle^2 (J_{Q_0} - J_q)\left(J_{Q_0} - \frac{1}{2}(J_{Q_0+q} - J_{Q_0-q})\right). \qquad [5]$$

where $J_q = \sum_{\vec{r}_n} J_n e^{-i\vec{q}\cdot\vec{r}}$ and $Q_0$ is the magnetic ordering wavevector. Since the powder averaged data presented in Figure 7 is within the first Brillouin zone, we can compare the dispersion curve directly. The low temperature powdered average magnetic spectral weight is presented in Figure 8. The high resolution ($E_i$ = 1.33 meV setting) is shown in Figure 8 (a) and the solid line is a plot of the $c$-axis dispersion with $J_{nn}$ set to 1.0 meV. The open circles represent the dispersion along the tetragonal $a$ direction. The solid lines provide a good description of the lower bound of the magnetic intensity with the open circles tracking the momentum dependence of the high-energy branch which extends beyond the dynamic range of the spectrometer. A lower resolution ($E_i$ = 3.3 meV setting) is shown in Figure 8 (b), illustrating a high-energy branch that extends up at least 1.5 meV. To obtain an estimate of the nearest neighbor exchange constant, we plot the peak position of this branch in energy as a function of $Q^2$ in Figure 8 (c). From the fit to a dispersion relation which is $\propto q^2$, we obtain an estimate of $J_{nn}$ = 1.0 meV. The fact that the curve does not intercept at $Q$ = 0 implies ordering a finite wave vector despite the dominant nearest neighbor ferromagnetic exchange constant, consistent with neutron diffraction data. Our simplified *2-J* model based on coupled chains provides a reasonable description of both the spin-wave dispersion (albeit in a powder sample), and also the magnetic ordering wavevector.

Based on the values for the exchange constants used to parametrize the spin excitations and the incommensurate wavevector, we can obtain an estimate for the Curie-Weiss temperature ($\Theta_{CW}$) using the formula,

$$k_B \Theta_{CW} = \frac{1}{3} S(S+1) \sum_n J_n \qquad [6]$$

$$= \frac{1}{3} s(s+1)(6 \times 1.0~meV + 4 \times -0.337~meV) \qquad [7]$$

$$\sim 67\,K \tag{8}$$

which is to be compared with an experimental value of 110 K[17]. The discrepancy may originate from the presence of a structural transition at $T_N$ which may alter the exchange constants. More accurate description of the magnetic Hamiltonian would be provided by a more complex exchange framework than the $J_{nn}$ - $J_{nnn}$ model considered here. However, this simple model in terms of two competing exchange constants does provide a means of understanding the spin fluctuations, the ordering wavevector, and an estimate of the Curie-Weiss constant.

These elastic and inelastic neutron scattering results discussed above shows that the nature of the magnetic phase transition, which was previously considered to be first order[1,40], is more complex than previously established. Figure 9 (a) shows the temperature dependence of $M(T)/M_0$ obtained from Rietveld fits of the neutron powder diffraction, and provides a value for the critical exponent $\beta$ of 0.14(1), which is similar to the values reported for $Li_2Mn_2O_4$[28] and $Pb_2Sr_2TbCu_3O_8$[56]. Both of these systems undergo similar transitions from two dimensional short range order to long ranged three dimensional order, but where the structural anisotropy from the layered systems creates Warren-type diffuse magnetic scattering[57]. $ZnCr_2Se_4$ has high cubic symmetry, so the formation of short ranged planes is not preferred within one particular plane, and so does not lead to substantial diffuse Warren scattering. As the magnetic transition does not evolve from a paramagnetic state, the magnetization increases faster on cooling than in classical three dimensional Heisenberg type magnet ($\beta \sim 0.36$).

The lack of ferromagnetic ordering at Curie-Weiss temperature of 110 K in $ZnCr_2Se_4$ is a result of the bond frustration and competing interaction. To evaluate the structural changes as a function of temperature, synchrotron X-ray diffraction experiments were performed between 80 – 500 K; the derived lattice parameters and structural parameter, u, are shown in Figure 9 (b). It can be seen that there is a gradual reduction of the rate of lattice contraction, such that below 100 K there is almost zero expansion. Indeed, previous studies have shown that $ZnCr_2Se_4$ enters a regime of negative thermal expansion between 75 and 21 K[16,23]. The structural parameter, u, that controls the bond angles within the $ZnSe_4$ tetrahedral unit and $Cr_4Se_4$ cube show little change, except a slight possible increase. The consequence of this is overall there is an increase in the Se – Cr – Se bond angle and subsequent weakening of the nearest neighbor ferromagnetic exchange and an increase to the components promoting antiferromagnetic exchange. Within this temperature regime there is a corresponding increase in the short range two-dimensional ferromagnetic order, which is overcome by the antiferromagnetic interactions to form a long ranged helical magnetic structure.

The manifestation of the spin-lattice coupling is very different in the case of $ZnCr_2Se_4$ compared with $ZnCr_2O_4$. The latter shows very strong geometric frustration with the large and negative Curie-Weiss constant implying strong antiferromagnetic exchange that is frustrated on a tetrahedral lattice. $ZnCr_2O_4$ shows a lowering of symmetry to $I\bar{4}m2$ at 12.5 K where long-ranged antiferromagnetic order is established, but the commensurate magnetic structure is associated with specific and identifiable distortions to the $Cr_4O_4$ cubes[26]. In the case of $ZnCr_2Se_4$, the bond frustration from competing interaction produces an incommensurate helical magnetic structure. The complex interactions between the magnetic ions will then be a function of the propagation vector along c. Thus each *ab* plane of $Cr^{3+}$ ions will experience a different combination of magnetic exchanges, which prohibits a specific and well-defined rearrangements of the $Cr_4Se_4$ cubes, and local static spatial distribution of Cr ions is stabilized. Further details of the distribution of Cr ions around its (0 0 0) position will required local structure techniques as diffraction averages these positions and only a three-fold increase of the thermal parameters is measured. However, maximum entropy reconstruction of the electron density map from synchrotron X-ray diffraction at 4 K, shown in Figure 10, does not indicate any significant static disorder.

## IV. Conclusions

We have synthesized a stoichiometric sample of the bond frustrated spinel, $ZnCr_2Se_4$, and shown that despite the presence of such frustration, the $Cr^{3+}$ ions order with a full moment. The magnetic ordering is accompanied by a distortion from the ideal cubic spinel to a tetragonal $I4_1/amd$ lattice. Previous reports of reduced ordered moments and more complex symmetry lowering is the result of non-stoichiometry that is prevalent in these systems. The frustration does, however, result in considerable local spin correlations at temperature around five times the long ranged ordering temperature, as measured by μSR spectroscopy.

## V. Acknowledgments

Funding for this project was provided by the Royal Society, EPSRC, and the Carnegie Trust for the Universities of Scotland. We acknowledge the support of the NIST, ESRF, ISIS and PSI for providing access to national research facilities used in this work.

**Figure and Table Captions**

**Figure 1** Synchrotron X-ray powder diffraction of $ZnCr_2Se_4$ showing (a) the observed, calculated and difference plot of the Rietveld refinement at 5 K and a comparison of a section of the pattern at (b) 50 K and (c) 5 K highlighting the tetragonal distortion to an $I4_1/amd$ space group at low temperature.

**Figure 2** Structural parameters obtained from fitting of the synchrotron powder X-ray diffraction data showing (a) both the Zn – Se and Cr – Se bond lengths are insensitive to changes with temperature, whereas there is a significant modification of their bond angles through the magnetoelastic transition at $T_N$ = 21 K and (b) the changes in the lattice parameters at the $Fd\bar{3}m$ → $I4_1/amd$ transition is accompanied by an increase in the Cr thermal factors.

**Figure 3** (a) Observed, calculated and difference plot of the powder neutron diffraction pattern of $ZnCr_2Se_4$ at 5 K at $\lambda$ = 2.078 Å using the BT1 diffractometer. The main magnetic reflections are marked with an asterisk. Inset highlights the low angle region, indexed in the tetragonal cell and (b) Observed, calculated and difference plot of the powder neutron diffraction pattern of $ZnCr_2Se_4$ at 1.5 K using -0.2 to 0.2 meV elastic contribution of the Disk Chopper Spectrometer (DCS) at $\lambda$ = 5.0 Å. Inset to (b) shows a graphical representation of the Cr moments within the spinel structure showing two dimensional planes of ferromagnetic order that propagate in a helical fashion down the *c* axis.

**Figure 4** Panel *a* Early relaxation times for ZF-µSR of $ZnCr_2Se_4$ showing very fast relaxation and small initial dip, and panel *b* example data of 0.3 T LF-µSR.

**Figure 5** Zero field measurement results of relaxation rates of the two components below $T_N$ and the value of the relaxation rate and power factor of the stretched exponent above $T_N$. Inset shows the temperature dependence of the fast relaxing component and the oscillation frequency.

**Figure 6** Fit results for the 0.3 T (3kG) longitudinal field measurement. At all temperatures two fractions are present with amplitudes $\lambda_1$ (fast) and $\lambda_2$ (slow). Above $T_N$ (from the region marked by the circle) they can be replaced by one stretched exponential form of relaxation.

**Figure 7** The temperature dependence of the powder averaged magnetic fluctuations taken with $E_i = 1.7$ meV on DCS. Panel *a* illustrates the spectrum in the magnetic ordered state at 2 K with b) showing that these are replaced by momentum and energy broadened excitations above $T_N$. c) plots a high temperature scan illustrating the background.

**Figure 8** (a) plots $E_i = 1.33$ meV data from DCS with the solid line and open points being the predicted single crystal dispersion along c and within the a - b plane. (b) illustrates a lower resolution $E_i = 3.3$ meV data illustrating a high energy branch extending up to 1.5 meV. (c) shows a plot of the peak position as a function of $Q^2$ illustrating a fit to a ferromagnetic dispersion curve ($\propto Q^2$).

**Figure 9** (a) Fit with numerous models to the evolution of magnetic moment in $ZnCr_2Se_4$ as a function of temperature, showing an unusual exponent of $\beta = 0.14(1)$, where the development of magnetic ordering occurs much quicker than expected within the Heisenberg model as a result of the short ranged ordering occurring below 100 K. (b) Evolution of the lattice and structural parameters in $ZnCr_2Se_4$ as a function of temperature.

**Figure 10** Reconstruction of the electron density of $ZnCr_2Se4$ from ID31 synchrotron powder X-ray diffraction data at 4 K using the maximum entropy method. No assumptions of symmetry are made during this analysis and the isotropic densities of all of the atoms present is indicative of little significant site disorder. Rietveld refinement of the neutron and synchrotron data shows the materials was stoichiometric as made.

**Table 1** Comparisons of the structural parameters obtained from Rietveld refinements of three tetragonal models of $ZnCr_2Se_4$ at 5 K using the ID31 diffractometer, showing $I4_1/amd$ as the true space group below the magnetic ordering temperature.

|  | $I4_1/amd$ (Model 1) | $I\bar{4}m2$ (Model 2) | $Fddd$ (Model 3) |
|---|---|---|---|
| $a$ (Å) | 7.41947(1) | 7.41946(1) | 10.49279(1) |
| $b$ (Å) | -- | -- | 10.49234(1) |
| $c$ (Å) | 10.48653(1) | 10.48653(1) | 10.48650(1) |
| $V$ (Å$^3$) | 577.267(1) | 577.267(1) | 1154.501(2) |
| $R_{exp}$ (%) | 11.84 | 11.84 | 11.84 |
| $R_{wp}$ (%) | 20.0 | 20.1 | 20.4 |
| $R_p$ (%) | 17.3 | 17.4 | 17.8 |
| $\chi^2$ | 2.86 | 2.89 | 2.97 |
| $R_B$ | 4.72 | 4.83 | 5.64 |
| $R_F$ | 4.28 | 4.38 | 5.72 |
| $U_{eq}$ Zn (Å$^2$) | 0.0062(2) | 0.0058(3) | 0.0067(7) |
| $U_{eq}$ Cr (Å$^2$) | 0.0059(2) | 0.0061(2) | 0.0067(7) |
| $U_{eq}$ Se (Å$^2$) | 0.0054(2) | 0.0053(5) | 0.0060(6) |
| $x$ Cr |  | 0.2488(8) |  |
| $z$ Cr |  | 0.3752(5) |  |
| $x$ Se1 | 0 | 0.7695(3) | 0.25951(8) |
| $y$ Se1 | 0.48066(3) | 0 | 0.2599(1) |
| $z$ Se1 | 0.25944(2) | 0.6157(2) | 0.25944(3) |
| $x$ Se2 | -- | 0.2694(3) | -- |
| $z$ Se2 | -- | 0.1350(2) | -- |

Table 1

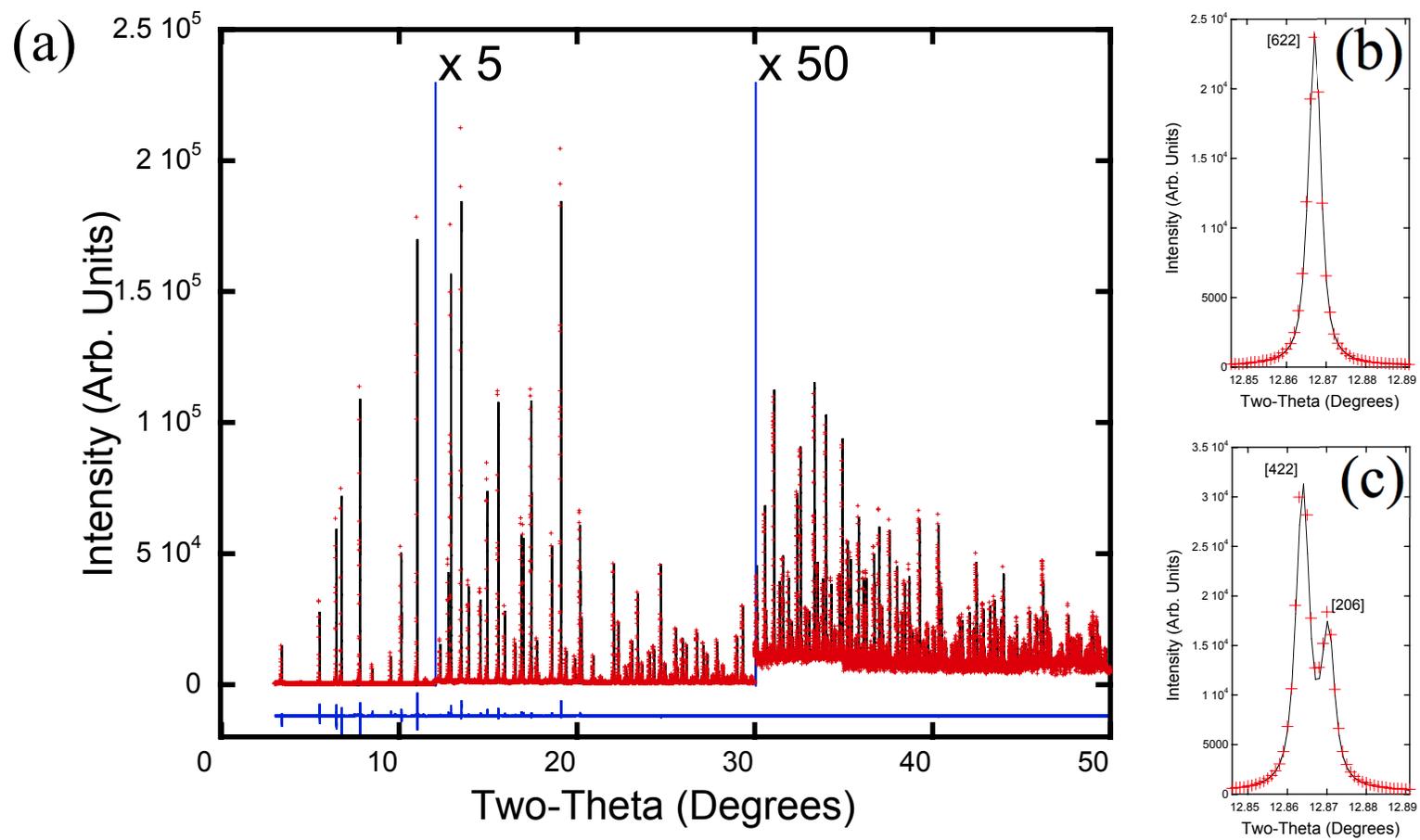

Figure 1

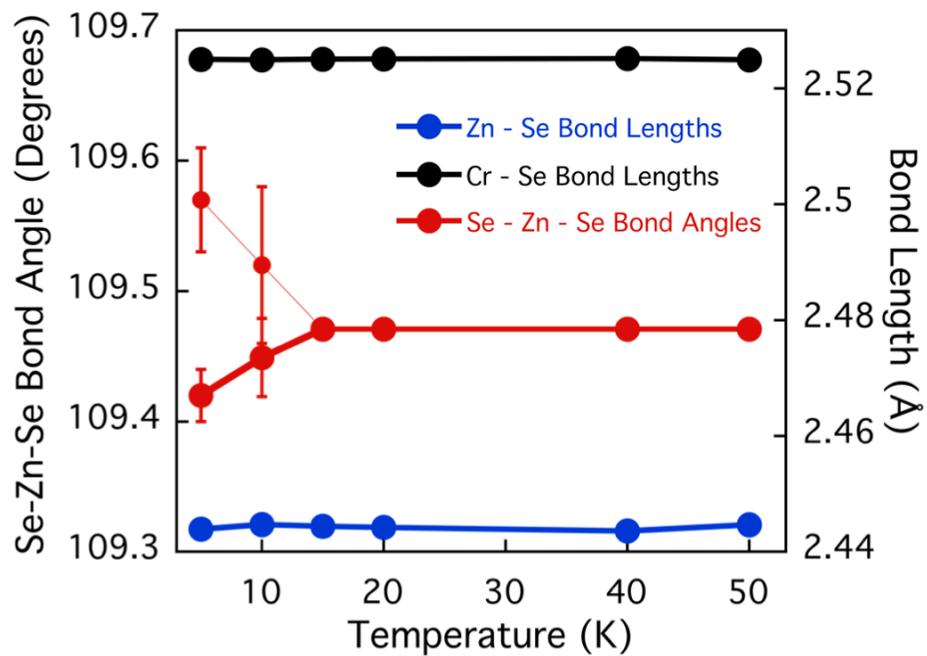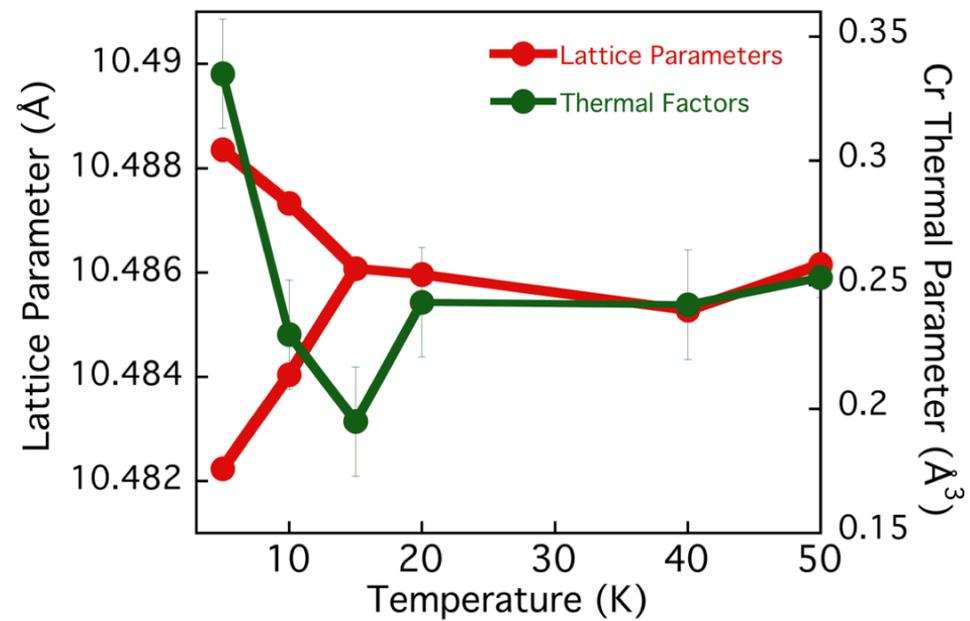

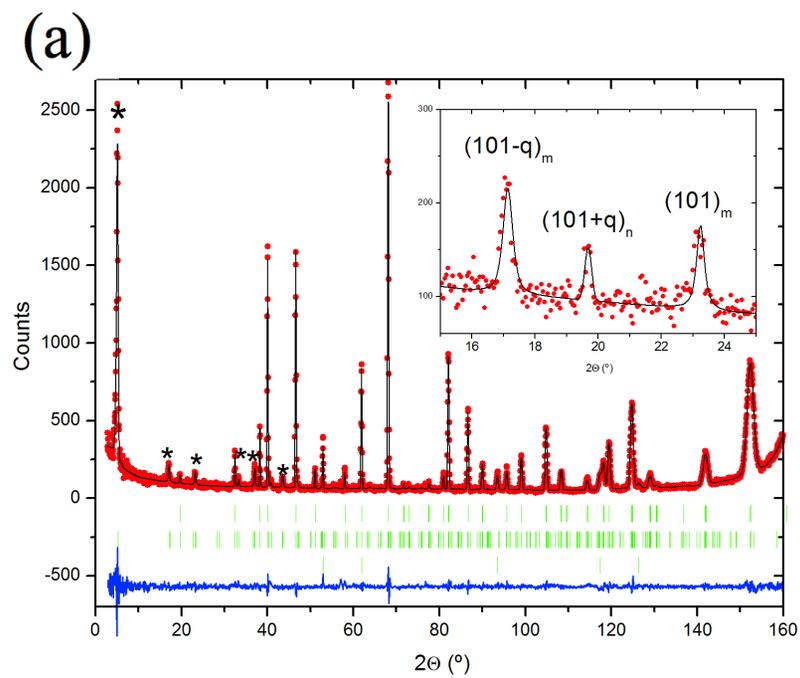
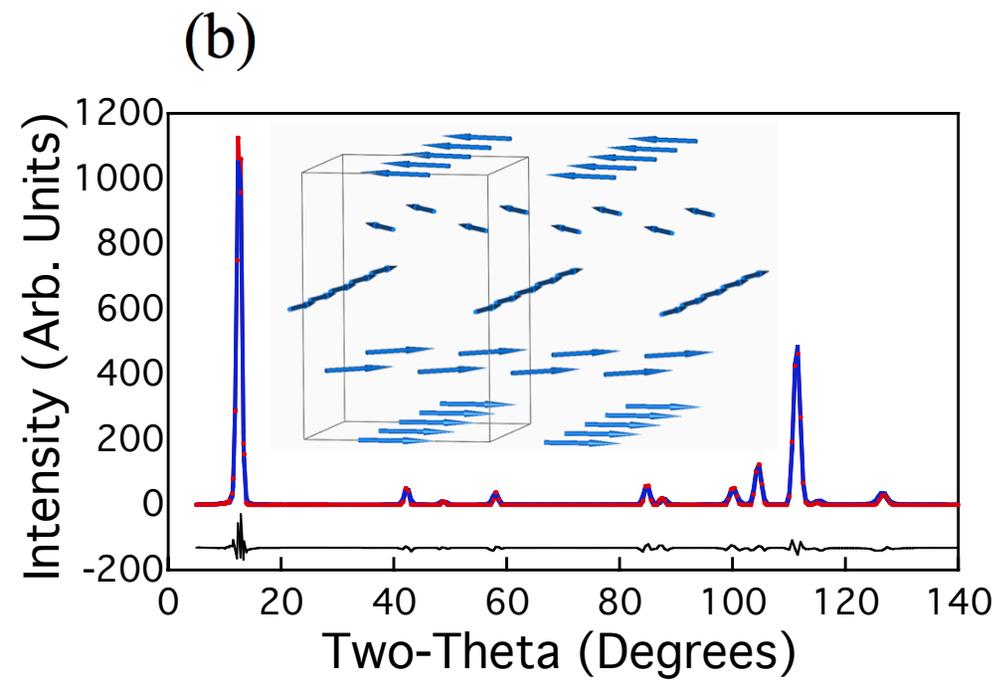

Figure 3

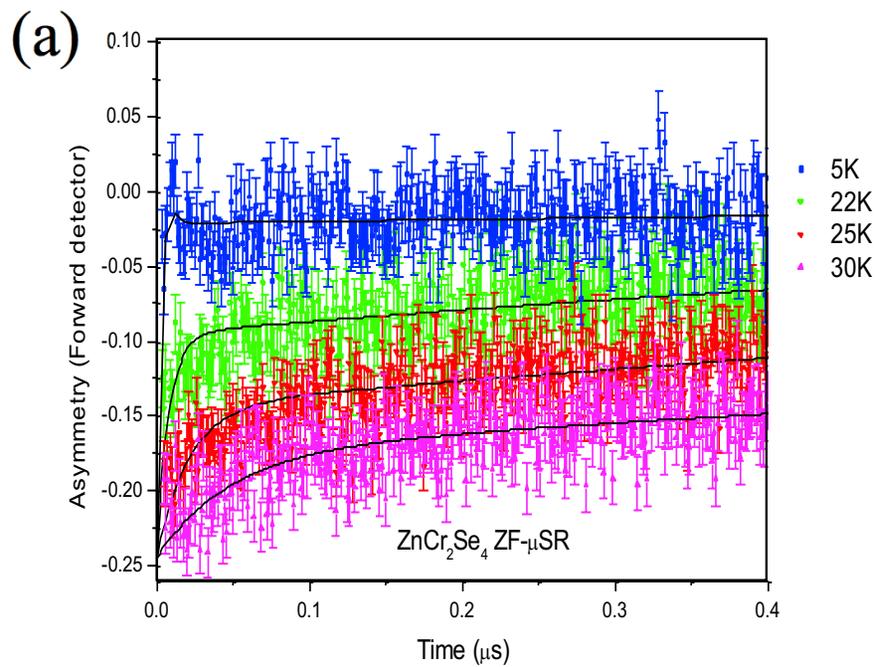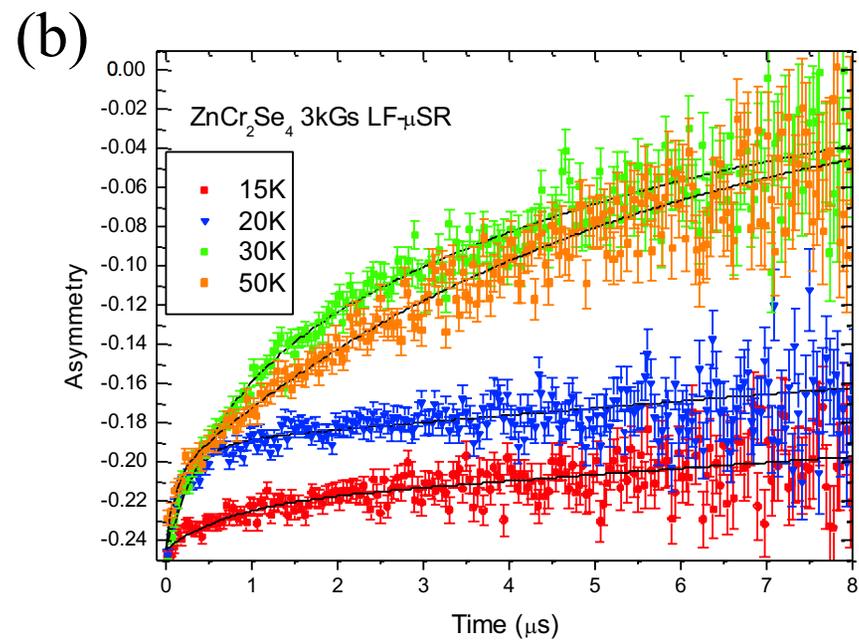

Figure 4

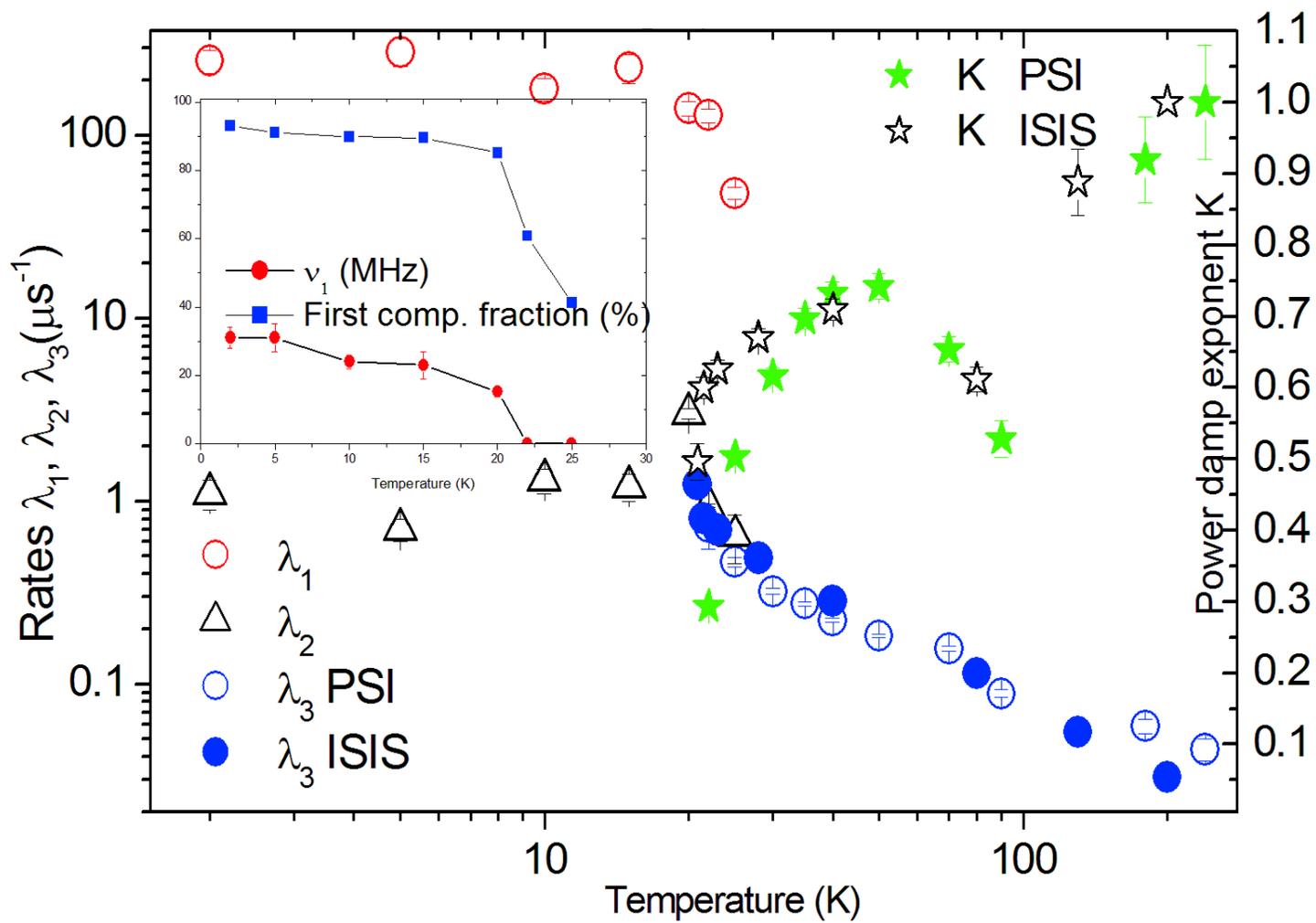

Figure 5

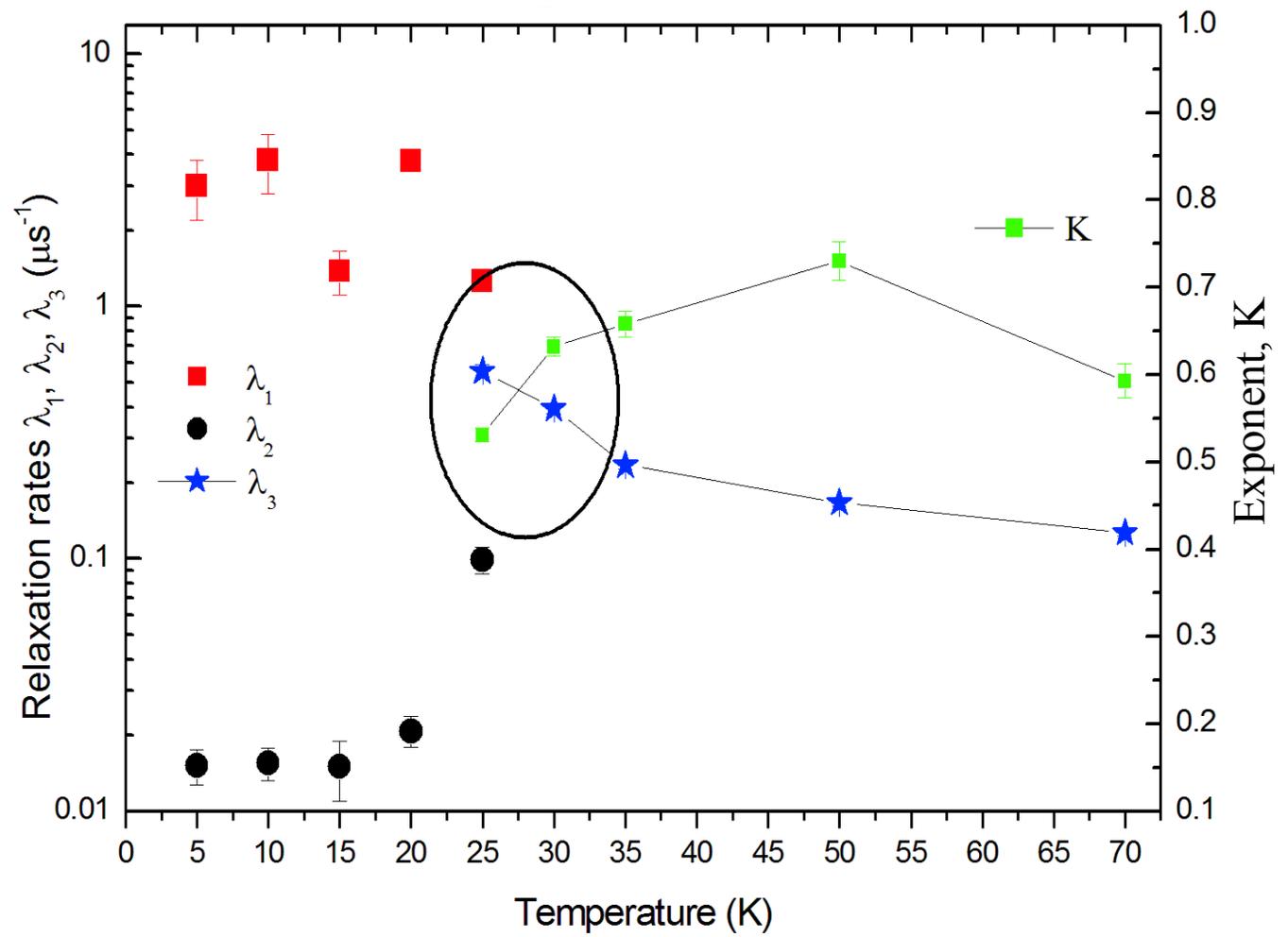

Figure 6

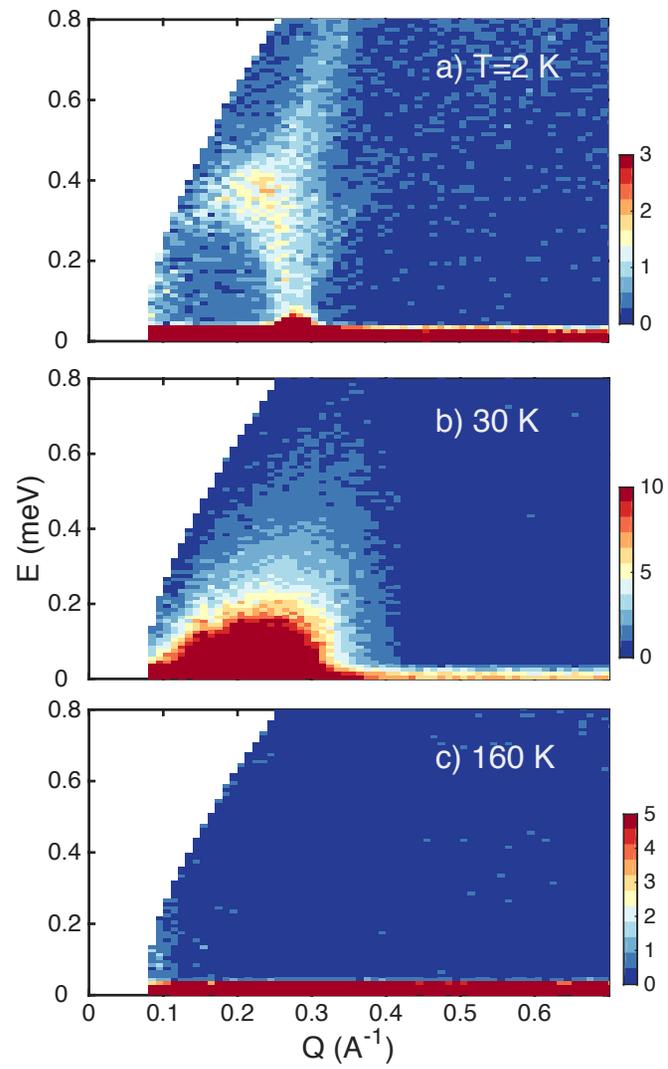

Figure 7

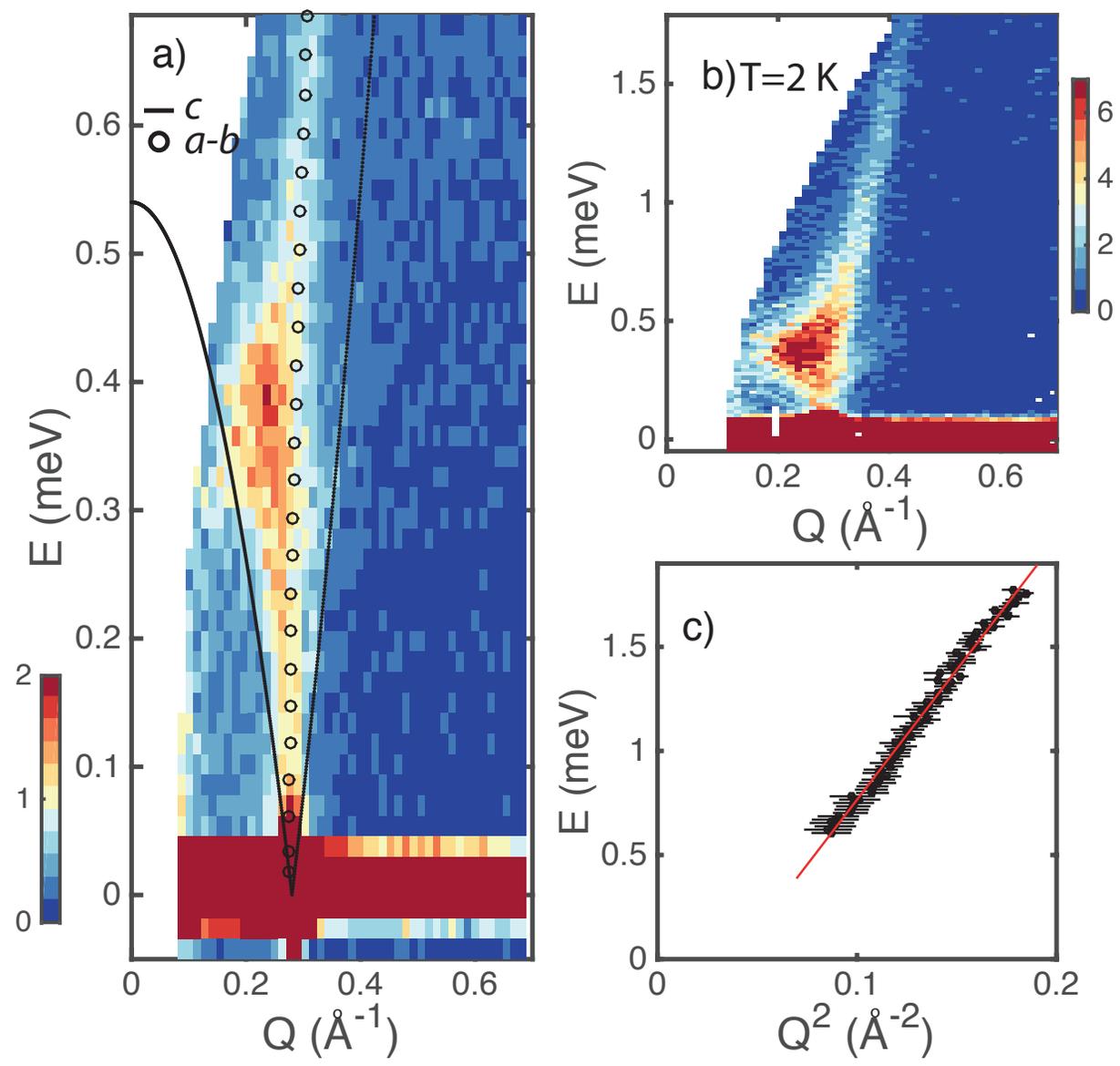

Figure 8

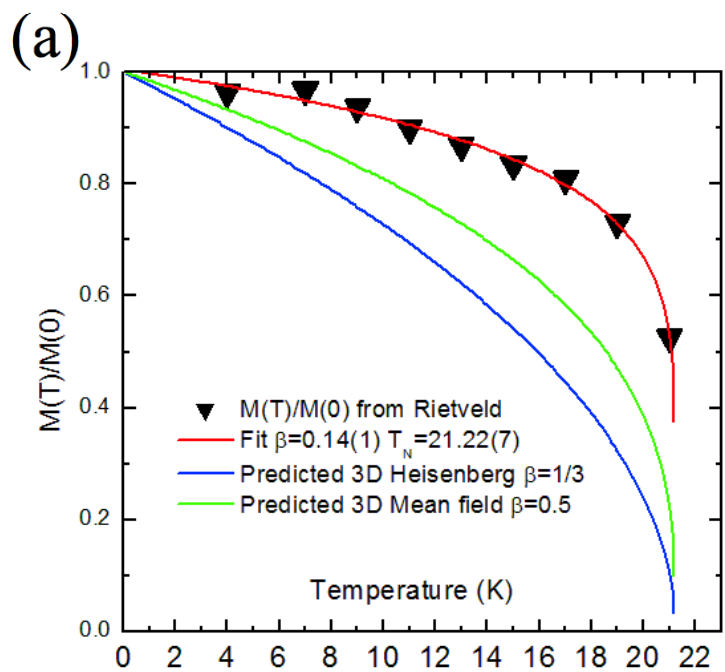# 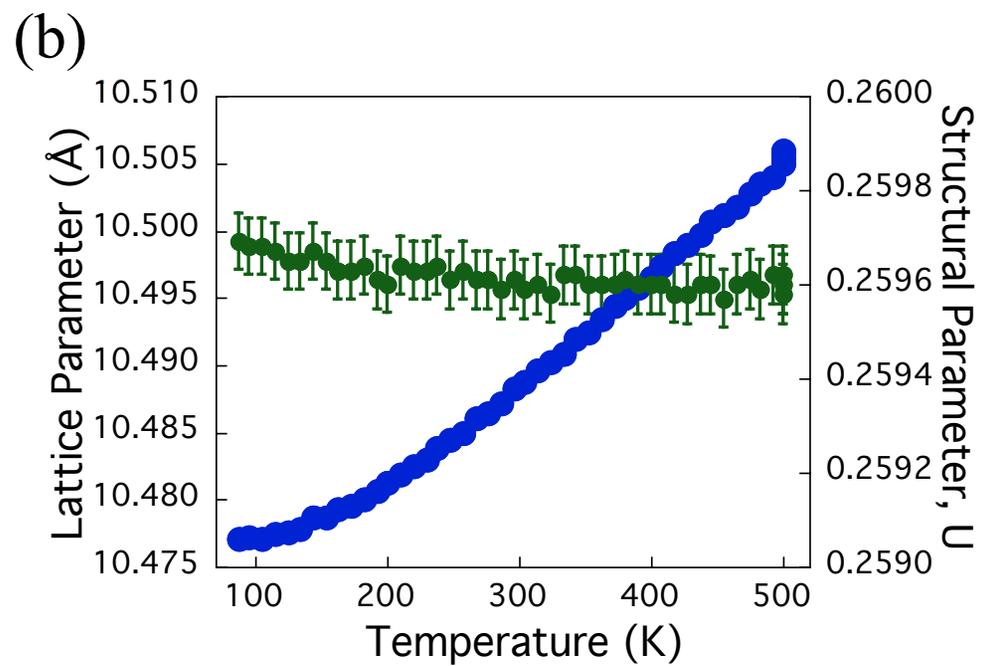

Figure 9

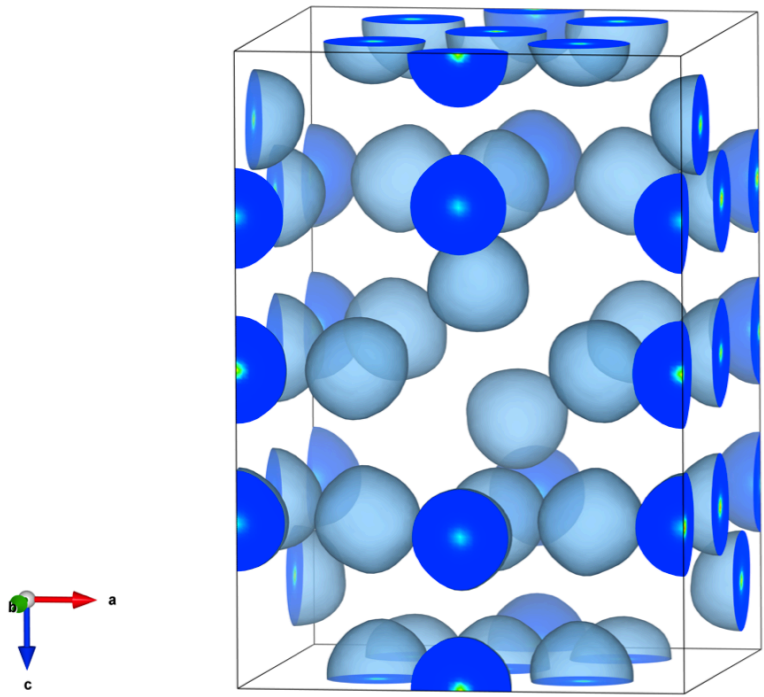

Figure 10